\documentclass[10pt, conference]{IEEEtran}
\usepackage[utf8]{inputenc}

\usepackage{amsmath,amsfonts,bm}

\def\eqref#1{equation~\ref{#1}}

\def\1{\bm{1}}

\DeclareMathAlphabet{\mathsfit}{\encodingdefault}{\sfdefault}{m}{sl}
\SetMathAlphabet{\mathsfit}{bold}{\encodingdefault}{\sfdefault}{bx}{n}

\usepackage{hyperref}
\usepackage{url}
\usepackage{soul}
\usepackage[normalem]{ulem}
\usepackage{xcolor}

\usepackage{xspace}
\usepackage{graphicx}
\usepackage{listings,multicol,tabularx}
\usepackage[caption=false]{subfig}  
\usepackage[export]{adjustbox}
\usepackage{amsmath}
\usepackage{ bbold }
\usepackage{textcomp}
\usepackage{listings}
\usepackage{cleveref}
\usepackage{algorithm}
\usepackage[noend]{algpseudocode}
\usepackage{tabularx}
\usepackage{multirow}
\usepackage{multicol}
\usepackage[noadjust]{cite}
\usepackage{booktabs}
\usepackage{balance}
\usepackage{enumitem}

\makeatletter
\newcommand{\linebreakand}{%
  \end{@IEEEauthorhalign}
  \hfill\mbox{}\par
  \mbox{}\hfill\begin{@IEEEauthorhalign}
}
\makeatother

\sethlcolor{yellow}

\newcommand\rmAP{\bgroup\markoverwith{\textcolor{blue}{\rule[0.5ex]{2pt}{0.4pt}}}\ULon}

\begin{document}

\author{
 \IEEEauthorblockN{Dmitry Pasechnyuk,\IEEEauthorrefmark{2} Anton Prazdnichnykh, Mikhail Evtikhiev,\IEEEauthorrefmark{1} Timofey Bryksin\IEEEauthorrefmark{1}}
    \IEEEauthorblockA{ \IEEEauthorrefmark{1}\textit{JetBrains Research},  \IEEEauthorrefmark{2}\textit{Mohamed bin Zayed University of Artificial Intelligence}}
    \IEEEauthorblockA{dmivilensky1@gmail.com, anton.prazdnichnykh@gmail.com, mikhail.evtikhiev@jetbrains.com, timofey.bryksin@jetbrains.com}
}

\title{Judging Adam: Studying the Performance of Optimization Methods on ML4SE Tasks}

\maketitle

\begin{abstract}
    Solving a problem with a deep learning model requires researchers to optimize the loss function with a certain optimization method. 
    The research community has developed more than a hundred different optimizers, yet there is scarce data on optimizer
    performance in various tasks.
    In particular, none of the benchmarks test the performance of optimizers on source code-related problems.
    However, existing benchmark data indicates that certain optimizers may be more efficient for particular domains.
    In this work, we test the performance of various optimizers on deep learning models for source code and find that the choice of an optimizer can have a significant impact on the model quality, with up to two-fold score differences between some of the relatively well-performing optimizers. 
    We also find that RAdam optimizer (and its modification with the Lookahead envelope) is the best optimizer that almost always performs well on the tasks we consider. 
    Our findings show a need for a more extensive study of the optimizers in code-related tasks, and indicate that the ML4SE community should consider using RAdam instead of Adam as the default optimizer for code-related deep learning tasks.
\end{abstract}

\section{Introduction}\label{sec:introduction}

Most of the deep learning algorithms involve the optimization of an objective function with a certain optimizer. 
Schmidt et al.~\cite{schmidt2021descending} collected a list of more than a hundred optimizers that can be used in deep learning, and also found that some optimizers perform better in particular domains. 
However, the majority of researchers and practitioners in the \textit{machine learning for software engineering} (ML4SE) domain use Adam optimizer~\cite{kingma2014adam}, usually with no particular argumentation for why this choice is optimal~\cite{ahmad2020transformer, peng2021integrating, wang2020learning, rabinovich2017abstract, GCNN, tranx, allamanis2017learning, guo2020graphcodebert, zuegner_code_transformer_2021}.
It is possible that this status quo is caused by the lack of information about the performance of optimizers in ML4SE tasks, as it is unclear how to infer the optimizer effectiveness for the ML4SE domain. 
This motivated us to study the efficiency of various optimizers on various ML4SE problems. 

In our work, we consider four models. \textsc{Code2Seq}~\cite{alon2018code2seq} and \textsc{TreeLSTM}~\cite{tai2015improved} are recurrent neural networks (RNNs) that use Abstract Syntax Trees (AST) together with the source code itself.
Both models are loosely based on the LSTM model, but \textsc{Code2Seq} derives a path-based representation of code from the AST, while \textsc{TreeLSTM} uses the AST information as is.
\textsc{CodeGNN}~\cite{leclair2020improved} is a graph neural network (GNN) that utilizes both sequential and AST representation of source code.
Finally, \textsc{CodeTransformer}~\cite{zuegner_code_transformer_2021} is an augmented transformer model for code, which combines information from multiple relations between tokens to calculate attention.

To test these models, we use two code-to-text generation problems as benchmarks: documentation generation and method name generation.
We chose these two ML4SE related problems to check the consistency of the optimizer performance. 
Moreover, these two problems are often used as the benchmarking problems for ML4SE models~\cite{lu2021codexglue, alon2018code2seq, leclair2020improved, zuegner_code_transformer_2021, fernandes2018structured, liu2022learning}.
For the documentation generation problem, we use the Python and Java parts of the CodeXGLUE dataset~\cite{lu2021codexglue} (\textsc{Java-CodeXGLUE, Python-CodexGLUE}), thus checking whether the programming language of the processed source code affects the optimizer performance.
For the method name generation problem, we also use two datasets: \textsc{Java-med}~\cite{alon2018code2seq}, and a smaller \textsc{Java-med-10per} dataset we created that allows us to check whether the effect of various optimizers depends on the size of the dataset.

In our study, we consider 24 optimization methods. 
Since running every combination of a model and an optimization method on every dataset is too computationally expensive, we use a two-step filtration scheme to study more promising optimizers in greater detail.
At the first step, we run all optimizer-model combinations on \textsc{Java-CodeXGLUE} to find six best-performing optimizers for each model.
At the second step, for each of these optimizers and the baseline Adam optimizer, we run the training on every remaining model-dataset combination.

Our results show that the choice of an optimization method can significantly affect the quality of the model predictions, with up to two-fold score difference. 
While the performance of various optimizers depends on the dataset and model, we find that RAdam optimizer (or its Lookahead modification) is the best optimizer out of the set of optimizers we consider and outperforms the default Adam optimizer in 14 model-dataset combinations out of 16.
The main contribution of this paper is thus the study of the effect of the optimizer choice on the model performance.
We also show how the optimizer performance depends on the choice of a model, an ML4SE task, and a dataset.
Finally, we formulate the preliminary recommendations for researchers and practitioners.

\section{Background}\label{sec:background}
A specific choice of an optimizer or corresponding hyperparameters can significantly change the quality of the model~\cite{schmidt2021descending}.
However, in the majority of the works we consider, researchers used Adam~\cite{kingma2014adam} as the default optimizer and did not report any attempts to tune its hyperparameters or choose another optimizer~\cite{ahmad2020transformer, peng2021integrating, wang2020learning, rabinovich2017abstract, GCNN, tranx, allamanis2017learning, zuegner_code_transformer_2021, guo2020graphcodebert}.
A rare exception is a paper by Hellendoorn et al.~\cite{hellendoorn2019global}, where the authors report tuning the initial learning rate for the model and the batch size.

\textbf{Benchmarking Optimizers.}
The literature on the optimizer performance is limited and does not consider the impact of optimizers on ML4SE problems.
The original papers that introduce new optimizers compare them only to a couple of well-known ones~\cite{zaheer2018adaptive}. 
There is also no standard way of carrying out the empirical evaluation in this case~\cite{schneider2019deepobs, sivaprasad2020optimizer}. 
Schneider et al.~\cite{schneider2019deepobs} suggest a DeepObs benchmark suite.
The benchmark has over twenty test problems from four application domains, which do not include ML4SE. 
It includes fully-connected networks, convolutional neural networks (CNN), and RNNs.
Metz et al.~\cite{metz2020using}, Choi et al.~\cite{choi2019empirical}, and Sivaprasad et al.~\cite{sivaprasad2020optimizer} suggest other datasets and show that a good choice of hyperparameters, such as $\beta_1, \beta_2$ for Adam, can improve the optimizer performance.

Schmidt et al.~\cite{schmidt2021descending} propose a benchmark of 15 most popular optimizers to find no optimizer to outperform others in all the tasks. 
However, they do find some optimizers to be better for certain architectures or tasks.
For example, RMSProp~\cite{tieleman2012lecture} performs best for natural language modeling tasks.
Similarly, it is reasonable to compare optimizer performance on ML4SE problems only if the problems, on which the optimizers were tested, are similar in scale, model architectures, and data format.

\textbf{ML4SE Specifics.}
The most popular architectures for ML4SE problems evolve with time, changing from RNN~\cite{tranx,alon2018code2seq} to Transformer~\cite{feng2020codebert}.
Sometimes, other architectures such as CNN~\cite{GCNN} or GNN~\cite{allamanis2017learning, leclair2020improved, hellendoorn2019global} are also used.
Even though Transformer and GNN architectures are widely used in ML4SE applications, they are hardly considered in the optimizer benchmarks mentioned above.  

Most ML4SE tasks use source code as either input data (\textit{e.g.}, commit message generation), output data (\textit{e.g.}, code generation), or both (\textit{e.g.}, code translation). 
It is possible to treat code as text and use NLP models such as CodeBERT~\cite{feng2020codebert} to process it.
However, compared to texts in natural languages, code has a richer structure that can be used to gain additional information. 
Thus, it is possible to extract data structures such as AST and use them to create better ML4SE models~\cite{alon2018code2seq, zuegner_code_transformer_2021, guo2020graphcodebert, allamanis2017learning}.

All these observations motivate us to study the performance of optimizers for ML4SE deep learning models.
Following the existing literature~\cite{lu2021codexglue, alon2018code2seq, leclair2020improved, zuegner_code_transformer_2021, fernandes2018structured, liu2022learning}, we study code-to-text problems of method name generation and documentation generation. 
We choose these two closely related problems to check the consistency of the optimizer performance.

\section{Methodology}\label{sec:internals}

In this paper, we aim to answer the following research questions:
\begin{itemize}[leftmargin=0.9cm]
    \item[$\textbf{RQ}_1$] Does the choice of an optimizer affect the model performance on ML4SE tasks?
    \item[$\textbf{RQ}_2$] How does the performance of optimizers on ML4SE tasks depend on the model architecture and the dataset?
\end{itemize}

\textbf{Tasks and Datasets.}
To evaluate the models, we use the problems of documentation generation (DG) and method name generation (MNG).
We use four different datasets in our experiments: \textsc{Python-CodeXGLUE} and \textsc{Java-CodeXGLUE}~\cite{lu2021codexglue} for DG, \textsc{Java-med}~\cite{alon2018code2seq} and \textsc{Java-med-10per} for MNG. 
We generate either up to 15 words or the first line of the documentaion (whatever is shorter) for DG, and up to 6 tokens for MNG. Reference values are truncated accordingly.
For the MNG task, we tokenize method names according to the \textit{CamelCase} convention.

We also subsample a new dataset \textsc{Java-med-10per} from \textsc{Java-med} to check whether the size of the training part of the dataset affects the optimizer performance. 
The training part of the new dataset is 10\% of the original dataset, the validation part is 50\%, and the test part is identical to the one of \textsc{Java-med}.
We subsample twice to get two different versions of \textsc{Java-med-10per} and find that the model performance does not depend on the subsampling, so in further experiments, we use only one of these versions.

To assess the trained models, we use F1 score for the method name generation task and chrF~\cite{popovic2015chrf} for the documentation generation task.
We use raw metrics scores to assess the optimizer impact and answer RQ1, and use the optimizer ranking (\textit{i.e.} which optimizer is best, second-best, etc. for every model-dataset combination) to answer RQ2.

\textbf{Models.}
We consider four models in our study that cover different architectures and use-cases. \textsc{Code2Seq} by Alon et al.~\cite{alon2018code2seq} is an RNN that encodes input source code using paths between terminal vertices of the AST of a program. 
Following Alon et al., we use a learning rate of $10^{-2}$ with a $0.95$ exponential decay.

\textsc{TreeLSTM} by Tai et al.~\cite{tai2015improved} is an extension of LSTM for the tree-structured data. 
Similarly to \textsc{Code2seq}, in our setup, we use a $10^{-2}$ learning rate with a $0.95$ exponential decay.

\textsc{CodeGNN} by LeClair et al.~\cite{leclair2020improved} is a GNN that uses both sequential and AST representation of a source code. 
Similarly to \textsc{Code2seq}, in our setup, we use a $10^{-2}$ learning rate with a $0.95$ exponential decay, and set the number of hops to 5.

\textsc{CodeTransformer} by Zuegner et al.~\cite{zuegner_code_transformer_2021} is an attention-based model that calculates attention scores from both program text and its structure representation.
Following Zuegner et al., we use a learning rate of $10^{-3}$ with a $0.95$ exponential decay.

\textbf{Optimizers}. 
The set of optimizers to consider is constructed as follows.
First, we take standard methods implemented in \texttt{pytorch}: SGD, Momentum, Adam, Adamax, and RMSProp~\cite{robbins1951stochastic, polyak1964some, kingma2014adam, tieleman2012lecture}.
Then, we add some of the optimizers implemented in the \texttt{pytorch-optimizer} package~\cite{Novik_torchoptimizers}. 
From \texttt{pytorch-optimizer}, we consider 10 top-cited optimization methods and the methods from repositories with more than 1,000 stars on GitHub.
The top-cited methods are SGDW, RAdam, AdaBound, Lookahead, Lamb, SWATS, Adafactor, Yogi, AdaBelief, DiffGrad~\cite{loshchilov2017decoupled, liu2019variance, luo2019adabound, zhang2019lookahead, you2019lamb, keskar2017improving, shazeer2018adafactor, reddi2018yogi, zhuang2020adabelief, dubey2019diffgrad}, and considering GitHub repositories with more than 1,000 stars adds the Ranger~\cite{wright2019ranger} method to this list.
We also consider Lookahead --- not a standalone optimizer, but a modification that can be applied to any optimizer.

Finally, we disregard SGDW and SWATS as a combination of SGD and Adam, and Ranger as a combination of Lookahead and RAdam.
Thus, the full list of tested methods is:  \textbf{Adam} (baseline method), SGD, Momentum, Adamax, RMSProp, RAdam, AdaBound, Lamb, Adafactor, Yogi, AdaBelief, and DiffGrad. 
Each of these methods is considered as a standalone method and in combination with Lookahead.

Following Roy et al.~\cite{roy2021reassessing}, we use F1 score and chrF~\cite{popovic2015chrf} to assess documentation generation models. 
In our study, we also considered BLEU, ROUGE, and METEOR metrics, but the results that we got with them were similar to what we get with the chrF metric scores.
Furthermore, unlike chrF, other metrics --- BLEU, ROUGE, and METEOR, --- depend on tokenization and there are numerous tuning options, which can all impact the comparison. For these reasons, we do not report the results of these metrics for the sake of brevity and clarity.

\textbf{Experimental Setup}.
Trying all models with all optimizers on all datasets would require many months of computation time. 
To make this pilot study possible, we use two conditions. 
First, we use standard hyperparameter values for the optimizers, such as $\beta_1 = 0.9$ for Adam. 
We also reduce the number of methods to test with the following filtration scheme. 

At Step 1, we use the \textsc{Java-CodeXGLUE} dataset, because it is small enough (181K methods as compared to 4M methods of \textsc{Java-med}), so we can test many methods on it.
Thus, at Step 1, we test all models with all optimizers on this dataset.

At Step 2, we use the results of Step 1 to select top-6 best-performing optimizers (plus the standard Adam) for each model to evaluate on the \textsc{Python-CodeXGLUE}, \textsc{Java-med}, and \textsc{Java-med-10per} datasets.

Since Lookahead is an optimizer modification, if both the original optimizer and its Lookahead modification belong to the top-6 optimizers, we replace the worse-performing optimizer of this pair with the next-ranking optimizer. 

In our experiments, we use the same random seed, and use a batch size of $512$. 
We do per-word tokenization for input and target sequences. 

\begin{table}[t]
    \caption{Optimizer performance on the Java dataset for documentation generation for all optimizers. 
    Six best-performing optimizers for each model that were selected for furhter experiments are highlighted in bold. 
    The reported metric is chrF.}
    \centering
    \label{tab:step1}
    \vspace{-0.2cm}
    \begin{tabular}{lc@{\hskip 0.1cm}cc@{\hskip 0.1cm}cc@{\hskip 0.1cm}cc@{\hskip 0.1cm}c}
    \toprule
    \multicolumn{1}{c}{\multirow{2}{*}{\textbf{Optimizer}}}      & \multicolumn{2}{c}{\textbf{Code2Seq}}  & \multicolumn{2}{c}{\textbf{TreeLSTM}}   & \multicolumn{2}{c}{\textbf{CodeGNN}}   & \multicolumn{2}{c}{\textbf{CT}}\\
    \cmidrule(lr){2-3}\cmidrule(lr){4-5}\cmidrule(lr){6-7}\cmidrule(lr){8-9}
    & \textbf{chrF} & \textbf{rank} & \textbf{chrF} & \textbf{rank} & \textbf{chrF} & \textbf{rank} & \textbf{chrF} & \textbf{rank} \\
    \midrule
    AdaBelief   & 19.8  & & 15.1 & & 22.9 & & 11.6 & \\
    LaAdaBelief & 21.1  & & 20.8 & & \textbf{24.3} & 4 & 8.9 & \\
    AdaBound    & 15.2  & & \textbf{21.0} & 7 & 21.8 & & 12.1 & \\
    LaAdaBound  & 14.2  & & 14.2 & & 20.8 & & 6.5 & \\
    Adafactor   & \textbf{21.8} & 5 & \textbf{23.8} & 4 & \textbf{24.8} & 3 & 4.4  & \\
    LaAdafactor & 19.8  & & 19.3 & & 23.4 & & 5.6 & \\
    LaAdam      & \textbf{22.2} & 4 & 19.1 & & 23.7 & & \textbf{35.3} & 2 \\
    Adamax      & \textbf{22.4} & 3 & \textbf{21.8} & 6 & 24.0 &  & 34.1 & \\
    LaAdamax    & 21.5 &  & 18.4 & & \textbf{25.3} & 1 & \textbf{35.6} & 1 \\
    DiffGrad    & \textbf{22.5} & 2 & 14.0 & & \textbf{24.3} & 4 & \textbf{35.2} & 3 \\
    LaDiffGrad  & 21.8 &  & \textbf{24.1} & 2 & 23.1 & & 34.8 &  \\
    Lamb        & 18.5 & & 15.3 & & 22.4 & & \textbf{32.5} & 7\\
    LaLamb      & 17.2 & & 13.7 & & 13.2 & & 29.8 & \\
    Momentum    & 19.7 & & 20.0 & & 22.6 & & 12.8 & \\
    LaMomentum  & 18.6 & & 17.0 & & 20.1 & & 0.2 &  \\
    RAdam       & \textbf{22.9} & 1 & 24.0 &  & \textbf{25.3} & 1 & 32.9 & \\
    LaRAdam     & 21.8 & & \textbf{25.5} & 1 & 24.1 &  & \textbf{34.3} & 6 \\
    RMSprop     & 18.2  & & 15.1 & & 21.1 & & 7.0 & \\
    LaRMSprop   & 15.1 & & 8.7  & & 19.0 & & 14.0 & \\
    SGD         & 13.7 & & 19.8 & & 22.4 & & 0.2 &  \\
    LaSGD       & 11.1 & & 16.7 & & 19.7 & & 0.2 &  \\
    Yogi        & \textbf{21.8} & 5 & 20.1 & & \textbf{23.8} & 8 & 31.4 & \\
    LaYogi      & 21.2 & & \textbf{21.3} & 5 & 23.0 & & 28.8 & \\
    \midrule
    Adam        & 19.5 & 16 & 11.7 & 23 & 20.7 & 20 & \textbf{34.8} & 4 \\ 
    \bottomrule
    \end{tabular}
    \vspace{-0.2cm}
\end{table}

We use one-sided signed-rank Wilcoxon test~\cite{wilcoxon1992individual} to check if we can reject $H_0: \text{score}(\text{method}_1) < \text{score}(\text{method}_2)$ or $H_1: \text{score}(\text{method}_1) > \text{score}(\text{method}_2)$. 
We choose Wilcoxon test, because we use metrics that are independently computed on each dataset sample and thus we have two distributions of metric scores originating from the input data.
To reformulate that, running the same model with different optimizers is essentially similar to testing on two dependent population samples with no prior for the distribution of the measured quantity inside each of the samples, and the Wilcoxon test is applicable in this setting.
We approve the ordering $\text{method}_1$-$\text{method}_2$ only if we can reject $H_1$ and cannot reject $H_0$ (with a $p$-value\;$=0.05$).
\begin{table}[hbt!]
    \caption{Optimizer performance on the Python dataset for documentation generation, on a subset of best-performing optimizers for each model. 
    3 best-performing optimizers for each model are highlighted in bold.} 
    \vspace{-0.2cm}
    \centering
    \label{tab:pyxglue}
    \begin{tabular}{lc@{\hskip 0.1cm}cc@{\hskip 0.1cm}cc@{\hskip 0.1cm}cc@{\hskip 0.1cm}c}
    \toprule
    \multicolumn{1}{c}{\multirow{2}{*}{\textbf{Optimizer}}}      & \multicolumn{2}{c}{\textbf{Code2Seq}}  & \multicolumn{2}{c}{\textbf{TreeLSTM}}   & \multicolumn{2}{c}{\textbf{CodeGNN}}   & \multicolumn{2}{c}{\textbf{CT}}\\
    \cmidrule(lr){2-3}\cmidrule(lr){4-5}\cmidrule(lr){6-7}\cmidrule(lr){8-9}
    & \textbf{chrF} & \textbf{rank} & \textbf{chrF} & \textbf{rank} & \textbf{chrF} & \textbf{rank} & \textbf{chrF} & \textbf{rank} \\
    \midrule
    RAdam       & 17.9 & & \multicolumn{2}{c}{----------} & \textbf{21.8} & 2 & \multicolumn{2}{c}{----------} \\ 
    LaRAdam     & \multicolumn{2}{c}{----------} & \textbf{21.0} & 2 & \multicolumn{2}{c}{----------} & \textbf{16.1} & 2 \\ 
    DiffGrad    & \textbf{18.4} & 2 &\multicolumn{2}{c}{----------} & 19.5 & & \textbf{16.6} & 1 \\
    LaDiffGrad  & \multicolumn{2}{c}{----------} & \textbf{22.0} & 1 & \multicolumn{2}{c}{----------} & \multicolumn{2}{c}{----------}  \\ 
    Adamax      & \textbf{18.0} & 3 & 12.8 & & \multicolumn{2}{c}{----------} & \multicolumn{2}{c}{----------} \\ 
    LaAdamax   & \multicolumn{2}{c}{----------} & \multicolumn{2}{c}{----------} & 19.3 & & 12.9 & \\ 
    LaAdam      & \textbf{18.5} & 1 & \multicolumn{2}{c}{----------} & \multicolumn{2}{c}{----------} & \textbf{15.7} & 3 \\ 
    Yogi        & 16.9 & & \multicolumn{2}{c}{----------} & \textbf{21.1} & 3 & \multicolumn{2}{c}{----------} \\ 
    LaYogi      & \multicolumn{2}{c}{----------} & 17.4 & & \multicolumn{2}{c}{----------} & \multicolumn{2}{c}{----------} \\ 
    Lamb        & \multicolumn{2}{c}{----------} & \multicolumn{2}{c}{----------} & \multicolumn{2}{c}{----------} & 15.5 & \\
    Adafactor   & 17.5  & & \textbf{18.9} & 3 & 20.3& & \multicolumn{2}{c}{----------}  \\ 
    LaAdaBelief   & \multicolumn{2}{c}{----------} & \multicolumn{2}{c}{----------} & 16.4 & & \multicolumn{2}{c}{----------} \\ 
    Adabound   & \multicolumn{2}{c}{----------} & 10.1 & & \multicolumn{2}{c}{----------} & \multicolumn{2}{c}{----------} \\
    \midrule
    Adam        & 16.1 & 7 & 10.4 & 6 & \textbf{23.0} & 1 & 14.0 & 5  \\
    \bottomrule
    \end{tabular}
    
    \vspace{0.3cm}
    \caption{Optimizer performance on Java-med dataset for method name generation, on a subset of best-performing optimizers for each model. 
    3 best-performing optimizers for each model are highlighted in bold.}
    \vspace{-0.2cm}
    \centering
    \label{tab:jm}
    \begin{tabular}{lc@{\hskip 0.1cm}cc@{\hskip 0.1cm}cc@{\hskip 0.1cm}cc@{\hskip 0.1cm}c}
    \toprule
    \multicolumn{1}{c}{\multirow{2}{*}{\textbf{Optimizer}}}      & \multicolumn{2}{c}{\textbf{Code2Seq}}  & \multicolumn{2}{c}{\textbf{TreeLSTM}}   & \multicolumn{2}{c}{\textbf{CodeGNN}}   & \multicolumn{2}{c}{\textbf{CT}}   \\
    \cmidrule(lr){2-3}\cmidrule(lr){4-5}\cmidrule(lr){6-7}\cmidrule(lr){8-9}
    & \textbf{F1} & \textbf{rank} & \textbf{F1} & \textbf{rank} & \textbf{F1} & \textbf{rank} & \textbf{F1} & \textbf{rank} \\
                \midrule
    RAdam       & 45.0 & & \multicolumn{2}{c}{----------}  & \textbf{46.5} & 2 &\multicolumn{2}{c}{----------}  \\
    LaRAdam     & \multicolumn{2}{c}{----------}  & \textbf{45.6} & 2 & \multicolumn{2}{c}{----------}  & \textbf{54.5} & 3 \\ 
    DiffGrad    & \textbf{51.2} & 1 &    \multicolumn{2}{c}{----------}      & \textbf{46.6} & 1 & 49.6 & \\ 
    LaDiffGrad  &    \multicolumn{2}{c}{----------}     & 26.9 & &    \multicolumn{2}{c}{----------}     &   \multicolumn{2}{c}{----------}     \\
    Adamax      & \textbf{50.9} & 2 & \textbf{45.1} & 3 &    \multicolumn{2}{c}{----------}       &   \multicolumn{2}{c}{----------}        \\
    LaAdamax  &    \multicolumn{2}{c}{----------}     & \multicolumn{2}{c}{----------} &  46.1 & &  \textbf{56.4} & 1 \\
    LaAdam      & 47.5& & \multicolumn{2}{c}{----------} & \multicolumn{2}{c}{----------} & 51.7 & \\ 
    Yogi        & 50.0 & &    \multicolumn{2}{c}{----------}       & 45.4 & &  \multicolumn{2}{c}{----------} \\ 
    LaYogi      &    \multicolumn{2}{c}{----------}           & 26.6 & &      \multicolumn{2}{c}{----------}  &     \multicolumn{2}{c}{----------}     \\
    Lamb        & \multicolumn{2}{c}{----------}  & \multicolumn{2}{c}{----------} & \multicolumn{2}{c}{----------} & \textbf{56.2}& 2\\ 
    Adafactor   & \textbf{50.5} & 3 & \textbf{46.7} & 1 & 45.9 & & \multicolumn{2}{c}{----------}\\ 
    LaAdabelief   & \multicolumn{2}{c}{----------}   & \multicolumn{2}{c}{----------} & 42.2 & & \multicolumn{2}{c}{----------} \\ 
    Adabound & \multicolumn{2}{c}{----------}  & 41.5 & & \multicolumn{2}{c}{----------} & \multicolumn{2}{c}{----------}\\
    \midrule
    Adam        & 43.3 & 7 & 34.3 & 5 & \textbf{46.4} & 3 & 47.0 & 6 \\ 
    \bottomrule
    \end{tabular}
    
    \vspace{0.3cm}
    
    \caption{Optimizer performance on Java-med-10per dataset for method name generation, on a subset of best-performing optimizers for each model. 
    3 best-performing optimizers for each model are highlighted in bold.}
    \centering
    \label{tab:jm10}
    \vspace{-0.2cm}
    \begin{tabular}{lc@{\hskip 0.1cm}cc@{\hskip 0.1cm}cc@{\hskip 0.1cm}cc@{\hskip 0.1cm}c}
    \toprule
    \multicolumn{1}{c}{\multirow{2}{*}{\textbf{Optimizer}}}      & \multicolumn{2}{c}{\textbf{Code2Seq}}  & \multicolumn{2}{c}{\textbf{TreeLSTM}}   & \multicolumn{2}{c}{\textbf{CodeGNN}}   & \multicolumn{2}{c}{\textbf{CT}}   \\
    \cmidrule(lr){2-3}\cmidrule(lr){4-5}\cmidrule(lr){6-7}\cmidrule(lr){8-9}
    & \textbf{F1} & \textbf{rank} & \textbf{F1} & \textbf{rank} & \textbf{F1} & \textbf{rank} & \textbf{F1} & \textbf{rank} \\
                \midrule
    RAdam       & 44.3 &  &\multicolumn{2}{c}{----------}  & \textbf{42.7} & 1 & \multicolumn{2}{c}{----------} \\ 
    LaRAdam     & \multicolumn{2}{c}{----------}  & \textbf{44.0} & 1 &\multicolumn{2}{c}{----------}  & \textbf{56.0} & 1\\  
    DiffGrad    & \textbf{45.4}& 3 &    \multicolumn{2}{c}{----------}       & 41.0 & & 54.8 & \\
    LaDiffGrad  &    \multicolumn{2}{c}{----------}            & 26.9 & &     \multicolumn{2}{c}{----------}   &    \multicolumn{2}{c}{----------}     \\
    Adamax      & \textbf{47.9} & 1 & \textbf{39.1} & 3 &    \multicolumn{2}{c}{----------}       &    \multicolumn{2}{c}{----------}       \\ 
    LaAdamax    &   \multicolumn{2}{c}{----------}    &        \multicolumn{2}{c}{----------}     &     41.6 & & \textbf{55.0} & 3 \\
    LaAdam      & 42.1 & & \multicolumn{2}{c}{----------}  & \multicolumn{2}{c}{----------} & 54.1 & \\ 
    Yogi        & 46.1 &  &    \multicolumn{2}{c}{----------}       & 39.6 & &   \multicolumn{2}{c}{----------}\\
    LaYogi      &   \multicolumn{2}{c}{----------}           & 26.6 & &       \multicolumn{2}{c}{----------}     &       \multicolumn{2}{c}{----------}\\
    Lamb        & \multicolumn{2}{c}{----------}&  \multicolumn{2}{c}{----------} & \multicolumn{2}{c}{----------}   & 55.1 & 2 \\
    Adafactor   & \textbf{46.8} & 2 & \textbf{42.8}& 2  & \textbf{41.4} & 3 & \multicolumn{2}{c}{----------}\\
    LaAdabelief   & \multicolumn{2}{c}{----------}  & \multicolumn{2}{c}{----------} & 39.8 &  & \multicolumn{2}{c}{----------}\\ 
    Adabound & \multicolumn{2}{c}{----------}  & 21.8 & & \multicolumn{2}{c}{----------} & \multicolumn{2}{c}{----------}\\
    \midrule
    Adam        & 39.0 & 7 & 33.5 & 4 & \textbf{42.4} & 2 & \textbf{55.0} & 3\\
    \bottomrule
    \end{tabular}
    \vspace{-0.4cm}
\end{table}
\section{Results and Discussion}\label{sec:conclusion}
    
The results of Step 1 are presented in Table~\ref{tab:step1}.
We abbreviate \textsc{CodeTransformer} as CT and abbreviate \texttt{Lookahead + Optimizer X} combinations as \texttt{LaX}.
We present the results of Step 2 in Tables~\ref{tab:pyxglue},~\ref{tab:jm},~\ref{tab:jm10}, we highlight the top-3 optimizer scores in bold. 
As we select only the top-6 performing optimizers + Adam for each model, the dashes in these tables mean that the optimizer was not selected at Step 1.

\textbf{RQ1. Optimizer impact.} The results show that the choice of an optimizer can have a huge impact on the model performance.
On Step 1, for the \textsc{Java-CodeXGLUE} dataset, even if we consider only the top-6 optimizers by performance, the improvement in the model scores over the worse-performing model-optimizer pair can reach up to 20\% (LaRAdam and Adabound scores for the \textsc{TreeLSTM} model, Table~\ref{tab:step1}).

On Step 2, the difference in the model scores for top-6 optimizers varies across the models and datasets and spans from 10\% for the \textsc{Code2Seq} model on \textsc{Python-CodeXGLUE} (Yogi and LaAdam optimizers, Table~\ref{tab:pyxglue}) to more than a two-fold difference for \textsc{TreeLSTM} on \textsc{Python-CodeXGLUE} (Adam and LaDiffGrad optimizers, Table~\ref{tab:pyxglue}). 
If we compare the top-performing optimizer to the baseline choice, Adam, we find Adam to be the best-performing optimizer in only one case (the \textsc{CodeGNN} model on \textsc{Python-CodeXGLUE}, Table~\ref{tab:pyxglue}), and in some cases the model with the best-performing optimizer improves the Adam-optimized model scores by up to 100\% (\textsc{TreeLSTM} on \textsc{Python-CodeXGLUE} with LaDiffGrad optimizer, Table~\ref{tab:pyxglue}).

\textbf{RQ2. Optimizer ranking.} To compare the optimizer performance, we compute optimizer ranks across models and datasets and calculate the average optimizer rank across all tasks. 
For example, RAdam is the best optimizer for the \textsc{CodeGNN} model on \textsc{Java-Med-10per}, thus its rank is 1 for this particular task.
For the ranking purposes, we merge the Lookahead and non-Lookahead versions of the optimizers. 
We find that the best optimizers are RAdam/LaRAdam (mean rank 2.4), DiffGrad/LaDiffGrad (mean rank 2.8) and Adamax/LaAdamax (mean rank 3.4). The baseline Adam method is significantly behind with a mean rank of 5.8.

\textbf{Performance consistency.}
To study the performance consistency, we consider the change in ranking of optimizers between various datasets and models.
The ranking of optimizers across all datasets (and, thus, all tasks) is relatively stable. 
For example, for each of the datasets, with the exception of \textsc{Python-CodeXGLUE}, the top-3 optimizers are RAdam/LaRAdam, DiffGrad/LaDiffGrad, and Adamax/LaAdamax; for \textsc{Python-CodeXGLUE}, instead of Adamax/LaAdamax we get Adam/LaAdam in the third place. 
This supports the hypothesis that the ranking of optimizers does not depend on the dataset or ML4SE task, and validates our filtration scheme. 

The ranking of optimizers for different models is less stable. 
For example, there are 6 different optimizers, which belong to the top-3 optimizers for at least one model, and no optimizer belongs to the top-3 optimizers for every model.
This means that the applicability of a particular optimizer may depend on the choice of the model.

\textbf{Takeaways.} 
We have found that the optimizer rankings are consistent across the choice of the dataset, but the best choice of an optimizer may depend on a particular model.
Despite the previous observation, some of the optimizers generally perform better than the others. 
Based on these observations, we suggest that ML4SE researchers and practitioners should consider trying several optimizers to improve the model quality.
We also suggest replacing Adam with LaRAdam as the optimizer of the first choice, and use RAdam as the fallback option.
In the case of a larger computational budget, we suggest to also try using DiffGrad or LaDiffGrad and Adamax or LaAdamax optimizers.
\section{Conclusion and future plans}\label{sec:threats}

In this paper, we study the impact of the optimizer choice on the model performance in ML4SE tasks.
We consider two tasks (documentation generation and method name generation), 4 datasets (\textsc{Python-CodeXGLUE}, \textsc{Java-CodeXGLUE} for documentation generation, and \textsc{Java-med}, \textsc{Java-med-10per} for method name generation), 4 models (\textsc{Code2Seq}, \textsc{TreeLSTM}, \textsc{CodeGNN}, \textsc{CodeTransformer}), and 12 different optimizers both with and without the Lookahead modification.
We use a two-step filtration scheme: first, we choose top-6 optimizers for each model, considering their performance on the \textsc{Java-CodeXGLUE} dataset, and then we study these optimizers and the baseline Adam method on the remaining datasets.
We find that the optimizer's performance depends on the model and is similar for different datasets, and RAdam/LaRAdam works best among the optimizers we consider.

The setup of our study leads to the following threats to validity, which also provide possible future work  directions.

\textbf{Choice of optimizers.} Our choice of optimizers was motivated by the practitioners' experience.  
We would like to do a survey of the ML4SE practitioners to check our assumptions.

\textbf{Choice of problems.}
We have chosen method name prediction and documentation generation as the benchmark problems. 
While these tasks are somewhat representative for the ML4SE domain, the observed performance may not be indicative for other tasks.
An exhaustive option would be to consider all tasks from the CodeXGLUE~\cite{lu2021codexglue} dataset, but this may be too computationally expensive.
In future, we would like to extend our benchmark set to include other ML4SE problems, and our first priority would be to include code completion and clone detection as the problems, which are useful in practice and distinct from the problems we consider in this paper.

\textbf{Choice of hyperparameters.}
Our training budget at this stage did not allow tuning the hyperparameters to improve performance.
Schmidt et al.~\cite{schmidt2021descending} argue that simply comparing optimizers without tuning hyperparameters is a viable strategy as the first approach.
Later, we would like to introduce hyperparameter tuning to find the best optimizers settings.
In particular, this tuning should include tuning optimizer hyperparameters (such as $\beta_1, \beta_2$ for Adam), tuning the batch size, and tuning the learning rate.
Despite this potential for improvement, our results are still significant as we compared different optimizers with the default optimizer setup for various ML4SE models and found that this setup could be improved even without proper hyperparameter tuning.
It would also be interesting to study the impact of the initialization on the optimizer robustness and performance.

\textbf{Choice of models.}
We consider models of a similar size (circa 30M parameters). 
While it is unfeasible to study optimizers for models with billions of parameters, we want to check if the optimizer performance depends on the model size.

\section*{Acknowledgments}\label{sec:acks}
We are grateful to Egor Spirin for his help in running and debugging the models, and to Yaroslav Golubev for helping to make this paper look clean and coherent.

\bibliographystyle{IEEEtran}
\balance
\bibliography{IEEEabrv,paper}

\end{document}